# Signal reflection, quantum non-locality, and delayed choice experiments

Moses Fayngold

*Department of Physics, New Jersey Institute of Technology, Newark, NJ 07102*

   Quantum nonlocality which is conventionally invoked for describing a composite entangled system is shown here to be a possible important characteristic of a single quantum object. To this end, we analyze some interactions of a single photon released from Fabry-Perot resonator with environment. The split photon state with oppositely moving parts is shown to obey quantum nonlocality despite the sharp edges truncating each part.
  Photon's post-release reflection from a plane mirror is considered. The changing shape of the form during reflection contains moving discontinuities in electric and magnetic components of the pulse. They originate from preexisting edges of the form and move together, first away from and then back to the mirror. At the end of the process, the pulse restores its original shape, with electric component reversed. Altogether, the process demonstrates conservation of moving discontinuities.
   The considered experimental setup may be used for some new versions of a delayed choice experiment, with various options for insertion of detectors and the respective time delays. In all cases, the delayed insertion does not have any retroactive effect on the process.

**Introduction**

We consider the life of a photon capable to carry a signal but emitted into two opposite directions. For our purpose, it is sufficient to review the simplest case – a signal from a source in the lowest non-vacuum photon state. For an efficient signaling, the state must have sharply defined edges. This may be a photon from a Fabry-Perot resonator (Sec.1).

The photon evolution after its release is considered in Sec.2. A wave form with sharp edges cannot be described by the conventional time evolution operator. We use instead a novel technique introduced in [1] and applicable to a signal of arbitrary shape.

Sec.3 describes the pulse reflection from a mirror at normal incidence. The edges of the incident pulse "send their marks" into the inner region of the changing form during the process. This produces two co-moving inner discontinuities $\mathcal{D}$ – one for electric and one for magnetic field. In the middle of the process the pulse shrinks to one half of its original length, but its initial shape (with electric field reversed) is restored by the end.

In Sec. 4 we consider some new versions of delayed choice experiment using the described arrangement. In all cases, the statistics of observed outcomes will remain unaffected and the delayed insertion of detectors does not have any retroactive effect on the process.

**1. Photon in a Fabry-Perot resonator**

Suppose the pulse originates from a photon eigenstate in a Fabry-Perot resonator of length $a$. A photon state $\Psi(\mathbf{r}, t)$ has two components representing electric and magnetic field, respectively. In terms of potential $U$, our photon is within a potential box. In the absence of motion along the mirrors, the fields **E** and **B** form a standing wave (SW) with vectors **E** and **B** perpendicular to each other and to resonator's optical axis $x$, so we have $\Psi(\mathbf{r}, t) \to \Psi(x,t)$. Assume the linear polarization and orient the $z$ and $y$-axes along **E** and **B**, respectively. Then, simplifying notations to $E_z \equiv E$ and $B_y \equiv B$, we can write

$$\Psi(x,t) = \frac{1}{\sqrt{a}} \begin{pmatrix} E(x,t) \\ B(x,t) \end{pmatrix} \tag{1.1}$$

Assuming the resonator to be lossless and placing the origin at its left end, we have on the outside:

$$\begin{pmatrix} E(x,t) \\ B(x,t) \end{pmatrix} = \begin{pmatrix} 0 \\ 0 \end{pmatrix}, \quad x < 0 \text{ and } x > a \tag{1.2}$$

The corresponding boundary conditions for $E$ are

$$E(0) = E(a) = 0 \tag{1.3}$$

Then the Maxwell equation $\vec{\nabla} \times \mathbf{B} = \mu_0 \left( \mathbf{j} + \varepsilon_0 \frac{\partial \mathbf{E}}{\partial t} \right)$ with $\mathbf{j} = 0$ between the mirrors imposes the conditions on $B$:

$$\left. \frac{\partial B}{\partial x} \right|_{x \to 0} = \left. \frac{\partial B}{\partial x} \right|_{x \to a} = 0 \tag{1.4}$$

On the mirror surfaces the **B** itself is discontinuous, jumping from some finite value within resonator to zero outside due to the surface current with $\mathbf{j} \to \infty$. This is consistent with the ban on non-zero **B** in a superconductor whose surface constitutes ideally reflecting mirror.

Within resonator, we have the set of the photon eigenstates satisfying conditions (1.3,4):

$$\begin{pmatrix} E(x,t) \\ B(x,t) \end{pmatrix} = \frac{1}{\sqrt{a}} \begin{pmatrix} E_0 \sin k_n x \cdot \cos \omega_n t \\ B_0 \cos k_n x \cdot \sin \omega_n t \end{pmatrix}, \quad 0 \le x \le a, \quad k_n = \frac{\omega_n}{c} = \frac{\pi}{a} n, \quad n = 1, 2, \ldots \quad (1.5)$$

The $E$ and $B$ fields in (1.5) are phase-shifted by $\Delta \phi = \pi / 2$ relative to each other. The $B$-field with amplitude $B_0 = E_0 / c$ is zero when $E$ is at its maximum and vice versa (Fig.1). Hereafter we consider a one-photon state.

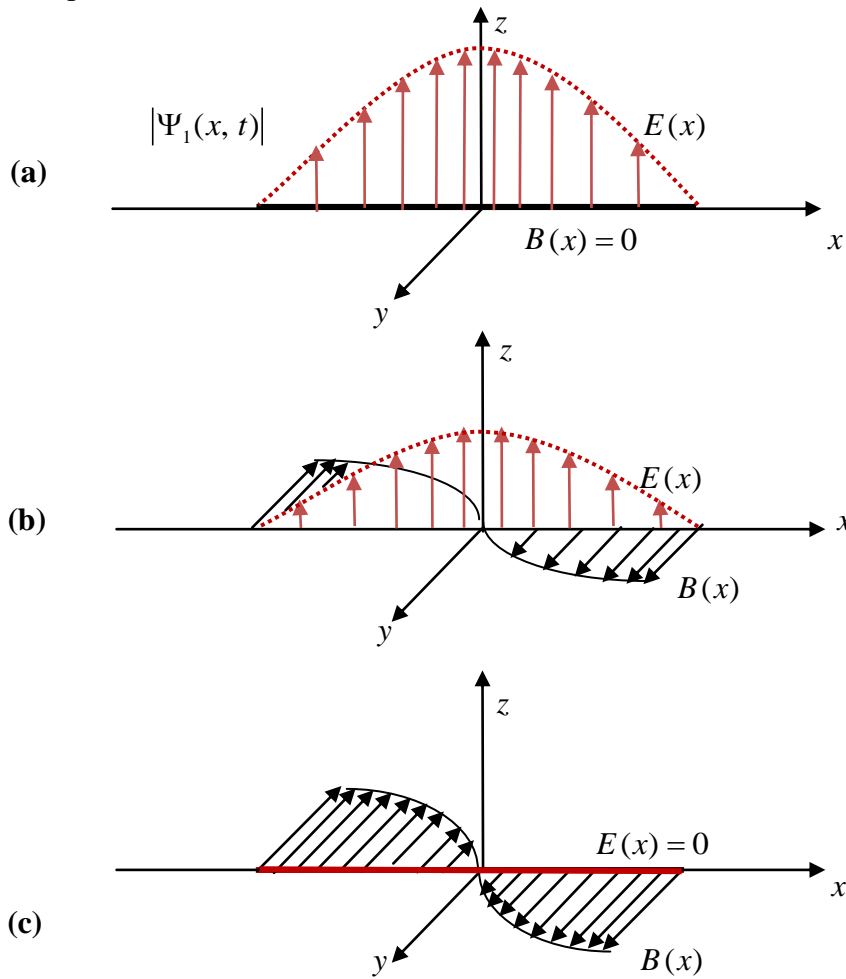

**Fig. 1**
Three snapshots from life of a trapped photon in its lowest non-vacuum state (not to scale)
(a) $t = 0$: $E$ is at its maximum, $B = 0$; (b) $t = (1/8)\,a/c$: Intermediate state;
(c) $t = (1/4)\,a/c$: $B$ is at its maximum, $E=0$

Expression (1.5) represents trivial oscillations with periodic conversions between the two fields described by ordinary solutions of the wave equation. Mathematically, it is analogous to a linear pendulum or mass on a spring with conversions between its potential and kinetic energies.

## 2. Release into free space

Let the resonator's mirrors get transparent at some moment taken as $t = 0$. This is equivalent to a sudden change of the Hamiltonian – the elimination of the box, leading to particle's escape. The state (1.5) is not an eigenstate of the new Hamiltonian, and it rapidly converts into two mutually receding pulses with $E(x,t) = c|B(x,t)|$ in each of them. The conversion is due to disappearance of the phase shift $\Delta\phi$ between the two fields. This happens because the boundary conditions at the edges disappear for both fields in synchrony with rapid disappearance of the surface currents; so eventually both fields become identically shaped, but the whole state splits into a pair of mutually receding wave packets of length $a$. The detailed process of reconfiguration of the trapped state into two mutually receding parts is interesting in its own right but its analysis is beyond the scope of this article.

It is now more appropriate to consider the resulting state as a continuous superposition of de Broglie's states with unlimited range of frequencies. Such superposition, while being virtual one before the release, becomes actual after it, describing two oppositely propagating pulses.

The evolution operator introduced in [1] for the forms with discontinuity leads to

$$E(x,t) = \sqrt{\frac{1}{2a}} \left\{ \begin{bmatrix} 0, & x - ct < 0 \\ \sin k_n(x-ct), & 0 \le x-ct \le a \\ 0, & x - ct > a \end{bmatrix} + \begin{bmatrix} 0, & x+ct < 0 \\ \sin k_n(x+ct), & 0 \le x+ct \le a \\ 0, & x + ct > a \end{bmatrix} \right\} \quad (2.1)$$

and similar for *B* but with the split parts having opposite signs

$$B(x,t) = \sqrt{\frac{1}{2a}} \left\{ \begin{bmatrix} 0, & x - ct < 0 \\ \sin k_n(x-ct), & 0 \le x-ct \le a \\ 0, & x - ct > a \end{bmatrix} - \begin{bmatrix} 0, & x+ct < 0 \\ \sin k_n(x+ct), & 0 \le x+ct \le a \\ 0, & x + ct > a \end{bmatrix} \right\} \quad (2.2)$$

Here we dropped the amplitudes $E_0$, $B_0$ to simplify the bulky expressions. The specific signs in (2.2) correspond to the left-handed coordinate system $(x, y, z)$ chosen in Fig.1.

A few resulting consecutive shapes of the whole state are shown in Fig.2. The initial state is a single packet ready to split, with $E(x,0)$ at its maximum and $B(x,0) = 0$. The intermediate state is two splitting but still partially overlapping wave packets. Finally we have two separate counter-propagating pulses retaining their respective shapes.

As seen from Fig.2, the resulting evolution of the state under the new Hamiltonian does not affect its polarization. Here and in all cases below, the *E*-filed remains in the $xz$ plane, whereas the *B*-field is in the $xy$-plane. The only changing factor is separation between the split parts, and then reshaping of the right form under reflection. The latter will be addressed in the next section.

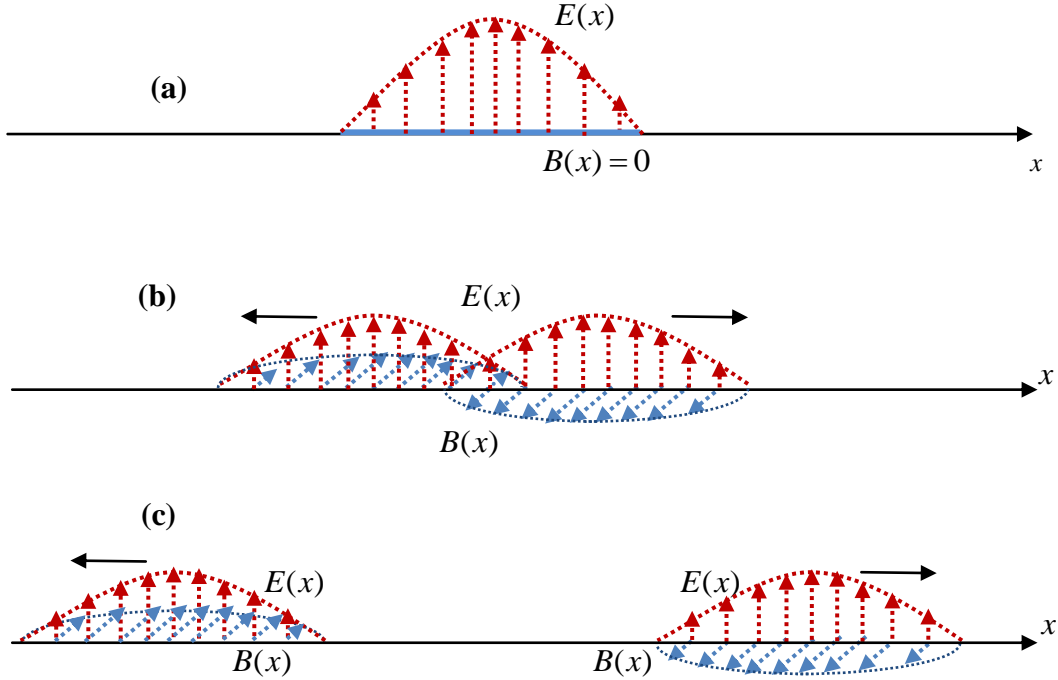

**Fig. 2**
Evolution of the photon released from potential box
Before the release, the photon was in the first eigenstate of the initial Hamiltonian. After release, it is in a superposition of eigenstates of the new Hamiltonian.
  (a) $t=0$;    (b) $0<t<a/2c$; the *net* $B(x)=0$ at the center of the overlap region;  actually shown are the *individual E* and *B* fields of the mutually receding pulses;    (c) $t>a/2c$

### 3. Reflection from a mirror

Reflection from a mirror is a textbook case for a monochromatic beam. But for a *finite-length* pulse considered here it has some important features that do not exist in the monochromatic case.

To simplify the analysis, we consider only one pulse from Fig.2 and treat it as a single object. Physically, this corresponds to opening only one (mirror-facing) side of resonator.

It is now convenient to shift the origin of the coordinate system. We assume the normal incidence from the left onto an ideally reflecting mirror M placed at $x=0$ (Fig.3). The mirror can be represented by the Hamiltonian allowing only one semi-space, so we will have $\Psi(x,t)=0$ for all $x>0$.

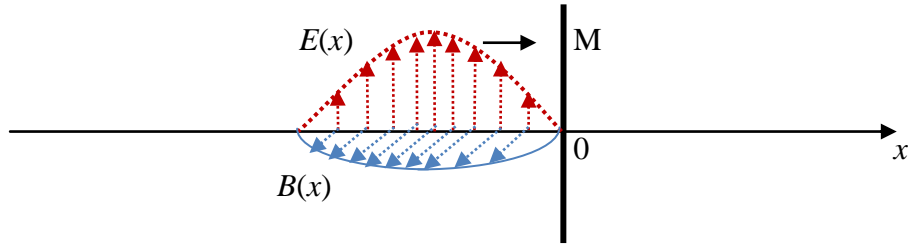

**Fig. 3**
The initial moment of an EM pulse reflection from a mirror

Another possibility of total reflection may be resonant scattering on a mono-atomic plane [2, 3]. Such planes are now intensely studied [4, 5], but they can totally reflect only a resonance-tuned monochromatic beam, which is not the case for a signal-carrying pulse.

Let the leading edge of the pulse reach the mirror at $t = 0$. At this moment, there appear corresponding boundary conditions for $E$ and $B$ similar to case (1.3, 4)

$$E|_{x=0} = 0 \quad \text{and} \quad \left.\frac{\partial B}{\partial x}\right|_{x=0} = 0 \qquad (3.1)$$

Conditions (3.1) hold during the time interval $\Delta t_a = a/c$. At each moment within this interval a certain fraction of the pulse is converted into a standing wave (SW). Consider one such moment $t_s \equiv s/c$ corresponding to propagation $s$ ($0 < s < a$) of the trailing edge (or to the equal back-propagation of the reflected leading edge). Then for $0 < s < a/2$ (the first stage of the process) the region $-s \leq x \leq 0$ will be occupied by SW for either field, while the trailing part of the incident pulse within $-a + s \leq x \leq -s$ remains the running wave (RW). For $a/2 < s < a$ (second stage), the SW will occupy the region $-a + s \leq x \leq 0$, while the RW will be the reflected leading part of the initial pulse in the region $-s \leq x \leq -a + s$. The $E$ and $B$ evolve differently due to difference in their boundary conditions; but both restore their initial shape (with $E$ reversed) at the end of the process.

For the initial state shown in Fig.3, an instant shape of the field at $t = t_s$ is described by

$$\Psi(x, t_s) = \frac{1}{\sqrt{a}} \begin{pmatrix} E(x, t_s) \\ B(x, t_s) \end{pmatrix}; \qquad (3.2)$$

$$\begin{pmatrix} E(x, t_s) \\ B(x, t_s) \end{pmatrix} = -\begin{pmatrix} \sin k_n (x - s) \\ \sin k_n (x - s) \end{pmatrix}; \qquad \begin{pmatrix} E(x, t_s) \\ B(x, t_s) \end{pmatrix} = 2\begin{pmatrix} -\sin k_n x \cdot \cos k_n s \\ \cos k_n x \cdot \sin k_n s \end{pmatrix} \qquad (3.3)$$

$$-a + s \leq x \leq -s \quad \text{(RW)} \qquad\qquad -s \leq x \leq 0 \quad \text{(SW)}$$

The corresponding probability density is:

$$\rho(x, s) = \frac{2}{a} \sin^2 k_n (x - s); \qquad \rho(x, s) = \frac{4}{a}\left(\sin^2 k_n x \cdot \cos^2 k_n s + \cos^2 k_n x \cdot \sin^2 k_n s\right) \qquad (3.4)$$

$$-a + s \leq x \leq -s, \quad \text{RW} \qquad\qquad -s \leq x \leq 0, \quad \text{SW}$$

Expressions (3.3) for either field match each other at $x_s \equiv -s = -ct_s$. But the corresponding *derivatives* are discontinuous, which is an imprint of the former leading edge. Starting at $t = 0$, this edge $x_s$ is receding from M at the speed $c$. When it meets with the trailing edge at a distance $s = a/2$ from M, the initial pulse length reduces by half, the RW domain shrinks to a point, and discontinuity $\mathcal{D}$ starts its way back to M.

In the second stage $t_s > a/2c$, the RW and SW-domains change to

$$-s \leq x \leq -a+s \quad \text{(RW)}; \qquad -a+s \leq x \leq 0 \quad \text{(SW)} \tag{3.5}$$

The RW are now reflected, that is $E(x,t) = \sin k_n(x+s)$ and $B(x,t) = -\sin k_n(x+s)$. Either field (but not their derivatives!) remains continuous at the border $x = -a+s$ between two domains. At the end we have completely reflected pulse detaching from the mirror, with reversed $E(x)$.

For illustration, consider five specific moments of the process in the simplest case $n=1$.
1) $s=0$. This is trivial – the arriving pulse just touches the mirror as shown in Fig.3. But at the same time it gives birth to the above-mentioned additional discontinuity *within* the pulse as will be seen in the next stages.
2) $s = a/4$. Then, denoting $(x/a) \equiv \xi$, we obtain from (3.3)

$$\begin{pmatrix} E(\xi,t_s) \\ B(\xi,t_s) \end{pmatrix} = -\begin{pmatrix} \sin \pi(\xi - 0.25) \\ \sin \pi(\xi - 0.25) \end{pmatrix}; \quad \begin{pmatrix} E(\xi,t_s) \\ B(\xi,t_s) \end{pmatrix} = \sqrt{2}\begin{pmatrix} -\sin \pi\xi \\ \cos \pi\xi \end{pmatrix} \tag{3.6}$$

$-0.75 \leq \xi \leq -0.25$ (RW-domain) $\qquad -0.25 \leq \xi \leq 0$ (SW-domain)

This is the middle of the first stage of reflection, with clearly seen inner discontinuities (Fig.4). They run away from the mirror.

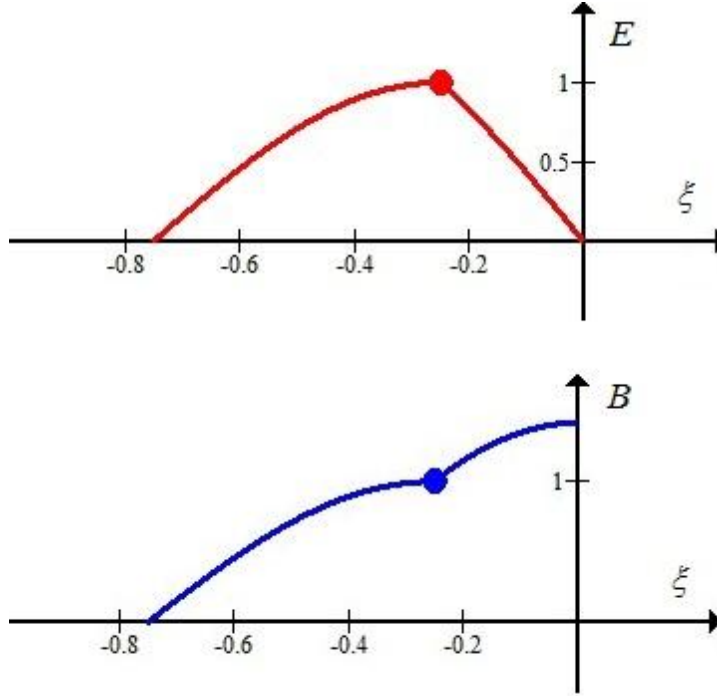

**Fig. 4**
Highlighted dots indicate discontinuities $\mathcal{D}$ in the respective derivatives

3) $s = a/2$.  $\begin{pmatrix} E(\xi,t_s) \\ B(\xi,t_s) \end{pmatrix} = \begin{pmatrix} 0 \\ 0 \end{pmatrix}$;  $\begin{pmatrix} E(\xi,t_s) \\ B(\xi,t_s) \end{pmatrix} = 2 \begin{pmatrix} 0 \\ \cos \pi\xi \end{pmatrix}$   (3.7)

$\xi = -0.5$ (RW-domain shrinks to a point)   $-0.5 \leq \xi \leq 0$ (SW-domain)

This is the middle of the whole process. The initial pulse is shrunk to one half of its length, and its trailing edge meets its former leading edge reflected from M. The $E$-field disappears and no room is left for RW at this moment (Fig.5).

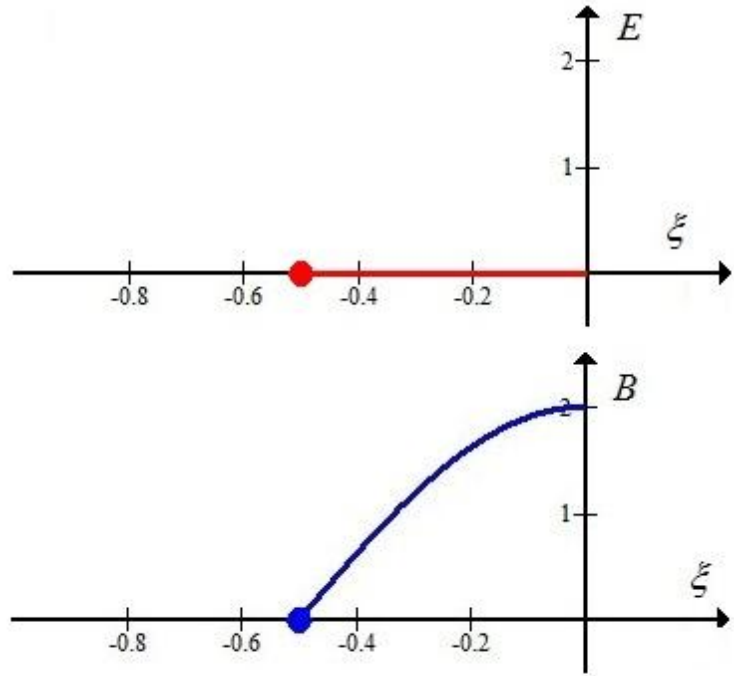

**Fig.5**

4) $s = (3/4)a$ (the middle of the second stage).

$\begin{pmatrix} E(\xi,t_s) \\ B(\xi,t_s) \end{pmatrix} = \begin{pmatrix} -\sin \pi(\xi + 3/4) \\ \sin \pi(\xi + 3/4) \end{pmatrix}$;  $\begin{pmatrix} E(\xi,t_s) \\ B(\xi,t_s) \end{pmatrix} = \sqrt{2} \begin{pmatrix} \sin \pi\xi \\ \cos \pi\xi \end{pmatrix}$   (3.8)

$-0.75 \leq \xi \leq -0.25$  (RW);   $-0.25 \leq \xi \leq 0$  (SW)

The inner discontinuities are now running back toward the mirror (Fig.6).

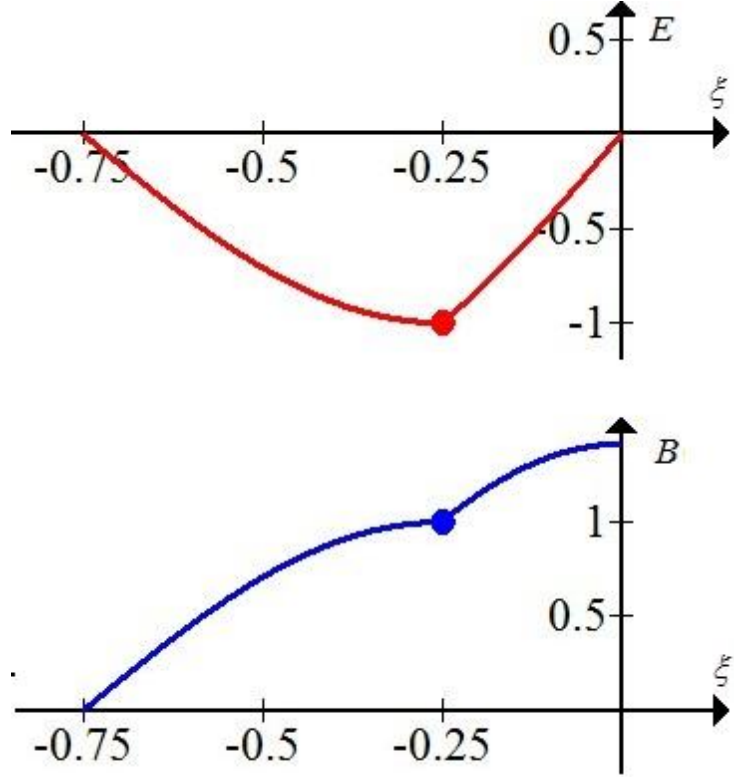

**Fig. 6**

5) $s = a$. This is trivial. The initial pulse is completely reflected and is de-touching from the mirror. Its shape differs from that in Fig. 3 only by reversed $E(x)$. The SW-domain disappears together with the inner discontinuities.

Consider now the energy distribution between the two fields in each domain. The reflected pulse energy in the RW domain (for $s \leq a/2$) is shared equally between the $E$ and $B$ fields and is obtained from (3.3) by the corresponding integration:

$$\mathcal{E}(\text{RW}) = 2 \int_{-a+s}^{-s} \sin^2 k(x-s)\, dx = a - 2s + \frac{\sin 4ks}{2k} \qquad (3.9)$$

At $s = a/2$ this fraction, together with its domain, dwindles to zero in accordance with (3.4).

In the SW domain, similar integration for the respective expressions from (3.3) gives the unequal energy distribution between the $E$ and $B$ fields as described by

$$\mathcal{E}_E = 2\left(s - \frac{\sin 2ks}{2k}\right)\cdot \cos^2 ks; \qquad \mathcal{E}_B = 2\left(s + \frac{\sin 2ks}{2k}\right)\cdot \sin^2 ks \qquad (3.10)$$

with the total in this region

$$\mathcal{E}(\text{SW}) = 2s - \frac{\sin 4ks}{2k} \qquad (3.11)$$

The sum of (3.9) and (3.11) multiplied by the normalizing factor $\left(1/\sqrt{a}\right)^2$ for both fields, gives, as it should, $\frac{1}{a}\left[\mathcal{E}(\text{RW}) + \mathcal{E}(\text{SW})\right] = 1$.

In the second stage $(s \geq a/2)$ of the reflection process, the RW domain hosts exclusively reflected wave $\begin{pmatrix} E \\ B \end{pmatrix} = \sin(x + ct)$ propagating away from the mirror. The same algorithm as before yields

$$\mathcal{E}(\text{RW}) = 2\int_{-s}^{-a+s} \sin^2 k(x+s)\,dx = 2s - a - \frac{\sin 4ks}{2k} \qquad (3.12)$$

For the SW domain we obtain

$$\mathcal{E}_E = \left(2(a-s) + \frac{\sin 2ks}{k}\right)\cdot \cos^2 ks; \qquad \mathcal{E}_B = \left(2(a-s) - \frac{\sin 2ks}{k}\right)\cdot \sin^2 ks \qquad (3.13)$$

$$\mathcal{E}(\text{SW}) = 2(a-s) + \frac{\sin 4ks}{2k} \qquad (3.14)$$

The sum $(1/a)\left[\mathcal{E}(\text{RW}) + \mathcal{E}(\text{SW})\right] = 1$, as before.

The most interesting feature of the process is the appearance of a moving discontinuity $\mathcal{D}$ in either field within the pulse, apart from the remaining discontinuities at the edges. Point $\mathcal{D}$ originates from the leading edge of the incident pulse and is moving from M to the left with the speed of light. In the second stage of the process $(s \geq a/2)$ the inner discontinuity originates from the former trailing edge and moves toward M. In either case, it acts as the "representative" of the respective edge in the initial pulse. The moving discontinuity within the signal carrier can be considered as an additional characteristic allowing reliable exchange of information even during the form interaction with environment. Extracting the imbedded information may be much more complicated in such cases, but it remains possible.

### 4. New versions of delayed-choice experiment

Here we discuss a few post-release rearrangements including insertions of some additional elements into the initial setup in Fig-s 1, 3. This may create new versions of the delayed-choice experiment (DCE). The initial versions proposed by Wheeler [6, 7] and intensely discussed later (see, e.g., [8-19]) were considering "which way decision" made by an emerging wave form; the decision included juxtaposition of one-way vs. two-ways propagation. The basic question was if and how the respective choices could be retroactively affected by future actions of the observer.

Most of the DCE were using quasi-monochromatic photon states. Such states are especially relevant when more than one way is available from the source to detector. Multiplicity invokes superposition and the respective interference pattern if at least two paths recombine, and in such cases narrow frequency range is necessary to produce observable pattern. Accordingly, the initial state must be a large pure ensemble of photons. A single photon will only make a single spot on observation screen or one detector click in a Mach-Zehnder interferometer.

One of the advantages of the suggested setup with Fabry-Perot resonator as a source for DCE is the possibility to record the exact time of the photon release (the moment when the mirrors are made transparent). This allows one to evaluate exactly the later moments or time intervals for planned rearrangements. Also, this permits using only one photon at each trial and broad frequency range associated with signaling. Accordingly, each trial with a single outcome is informative, e.g. in position measurement.

Using localized photon states from Fabry-Perot resonator invites a relevant discussion of the role of quantum non-locality in state (2.1, 2) and propagation-detection relationship in the observation process.

Quantum non-locality may enable a single particle even when emanated from a small region to "feel" practically instantly the whole environment no matter how remote. This is an important property of radiation from an initially localized system in a Gamow state [20, 21], e.g., photon emission from an excited atom or radioactive decay of an unstable particle.

Another manifestation of quantum non-locality is "freezing" of an excited state when placed in a non-hospitable cavity. If the cavity modes are detuned from an energy eigenstate of the particle to be emitted, the system freezes in its excited state even if utterly unstable when in a free space [22-25]. The time indeterminacy associated with a Gamow state makes the distinction between its past and future fuzzy and enables the photon to collect information about its not yet started life. The would-be photon learns about the cavity's unsuitability *before* getting its chance to materialize. This pre-creational knowledge precludes the creation.

The situation with a frozen excited state is in some respects similar to suppression of inelastic channels in the planar resonant scattering [2, 3] mentioned in the previous section. Even though the mechanism in the two cases is different, the outcome is the same.

Some interpretations of QM proclaim, as mentioned in the beginning of this section, somewhat opposite effect for a particle emitted into an open space: if it "feels" any possibility to get observed (and thereby probably absorbed!) in a future measurement, it selects the optimal way to actualize it. If the particle is already in a superposition of motions along the oppositely-directed paths L (left) and R (right)

$$|\Psi\rangle = \frac{1}{\sqrt{2}}\left(|L\rangle + e^{i\eta}|R\rangle\right) \qquad (4.1)$$

but the observer places a detector, say, in path L, the state (4.1) instantly collapses to $|L\rangle$ leading to detector.

The true meaning of the misleading term "collapse" has been discussed in [1, 26], where we had suggested "*Instant Reconfiguration*" as a far better term. We squeeze it here to "*InstReguration*" for brevity (abbreviation "IR" would be confusing with Infra-Red).

In the mentioned interpretations of QM, the *InstReguration* to detector-leading path occurs at the moment of inserting detector, which may be long *before* its being actually reached by the pulse; hence the term "retroactive behavior".

If now the detector is suddenly relocated from L to R, the particle can retroactively switch to state $|R\rangle$. And if we make L and R intercross by a system of mirrors and relocate the detector to the respective intersection, the particle would *InstRegurate* back to state (4.1).

According to this view, it looks like there is some intimate telepathic link between the observer and the object to be observed, with the object generally inclined to actualize observation.

The suggested experiments [6, 7] offered a way to test this interpretation (see e.g., [11]). The Wheeler's thought experiment tests the light from a distant quasar passing by the massive black hole between the quasar and the Earth. The hole acts as a gravitational lens. Let a detector D be placed between the quasar and the Earth. The photon makes the which-way decision (whether to fly on the right or left of the lens or take both ways) depending on D's location. If the D is placed in one of the paths the photon chooses this path. If D is later moved to the region of paths' intersection, the photon retroactively "corrects" its initial choice by switching to both ways. In this interpretation, the photon had been initially "fooled" to take only one way but then it triumphantly appears approaching along two different paths to interfere with itself.

But as emphasized in [10-19], the experiments did not show any retroactive effects in QM. And the "chosen path" view is simply refuted by passing the pulse, say, through a beam-splitter. If symmetric, it would split the pulse into an equally-weighted superposition of transmitted and reflected states. Their only difference from (4.1) is that the resulting beams make $90^o$ with each other. Putting a detector into one of the beams does not change their 50-50 statistics.

Now we discuss interrelations between quantum non-locality, delayed choice, and signaling.

Since the signal-carrier in Fig-s 1, 2 has the zero field outside the packet, non-locality can be quantified. It seems to consist of two separate and mutually receding domains of length $a$ each. But the actual situation is more subtle. The two "half-selves", albeit receding from each other, are parts of one-photon state. This makes communication between them instantaneous (without any signaling!) regardless of intermediate space; so quantum non-locality must still include all space between the pulses. Numerically, it increases with time until one of the pulses collides with another object (the mirror M in our case).

Usually, quantum non-locality is attributed to a composite system with spatially separated parts. A textbook example is a pair of two spin-entangled electrons A and B with the zero net spin. Then, if Alice measures, say, vertical spin component of A and finds it in a state $|\uparrow\rangle_A$, she immediately knows that Bob will find B in state $|\downarrow\rangle_B$, and vice versa.

Our system is a *single* photon, but it exhibits the similar features when in a split state. We can find upon measurement the whole photon either on the left or on the right from its source S. Ascribe the bit value 0 to the state of the photon not found in a measured pulse, and 1 when it is found in the pulse. Then if Lucy finds our photon on the left of S (state $|1\rangle_L$), she immediately knows that Ron will find $|0\rangle_R$ on the right. So a single particle demonstrates quantum non-locality with somewhat different "face", but in the same kind of measurements with correlated outcomes as a composite entangled system [1, 26].

We can define the split photon's non-locality range at a moment $t$ after release as the distance between its extreme edges:

$$S(t) = 2ct + a \qquad (4.2)$$

This is just the total range of the system (2.1, 2), with $S - a$ being an instant distance between the centers of the pulses. In experiment without mirror, the range (4.2) may increase unboundedly until the corresponding position measurement provoking both split parts to InstRegurate to one location.

With the mirror at a distance $D > a$ from the source, non-locality reaches its maximum

$$S_M = 2D + a \qquad (4.3)$$

at the end of reflection. The corresponding moment of time is

$$t_D \equiv (2D + a)/2c \qquad (4.4)$$

At $t > t_D$, both pulses move one way with a fixed distance $S_M - a = 2D$ between their centers.

In conventional QM, there is no need for a wave packet to "pre-program" its subsequent behavior depending on the shape of the world. Its evolution is universally determined by a wave equation and Born rule: a quantum object propagates as a wave and is recorded as a particle. In our experiment, the photon from source S takes both ways regardless of the shape of environment. Its subsequent behavior only reveals intrinsically probabilistic nature of QM.

Now we place the origin at the center of the source and consider some new versions of the delayed choice experiment.

**I**. The arrangement remains the same as before, and we perform in addition the photon position measurements in the region

$$D - a < x < D, \qquad (4.5)$$

within the corresponding time interval

$$t_D - a/c \leq t \leq t_D \qquad (4.6)$$

This would be equivalent to effective measurement of the amplitudes and the respective energy density distribution calculated in the previous section. It can be done, e.g., by shooting electrons into region (4.5) one at a time concurrently with the reflection process and recording locations of the resulting scattering interactions and their statistics (Fig.7). To simplify the discussion, we assume here that each shot leads to observable scattering. This part of the experiment may be difficult to perform since the range (4.5) is, as shown in the previous section, itself shrinking down to its half, and then restores its initial length by the end of reflection.

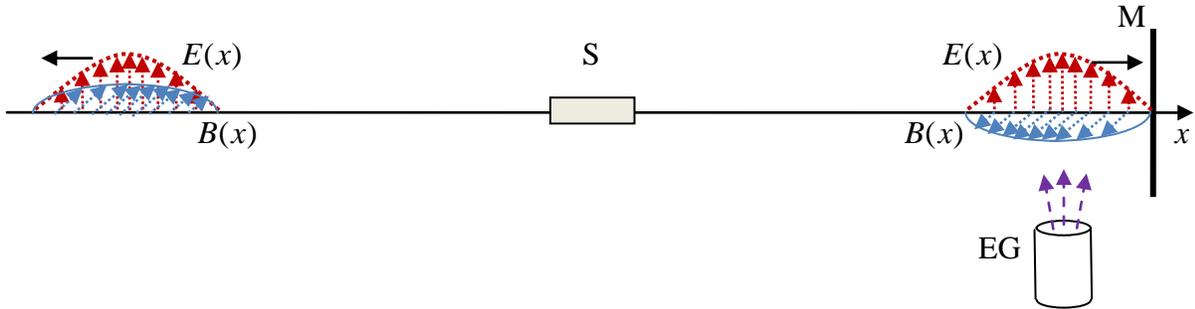

**Fig. 7**
The split photon from the source S at the moment when its right "half-self" touches the mirror
EG – an electron gun emitting electrons at a controlled rate and time

The setup in Fig.7 is a "reversed version" of the "which slit" measurement as described in [27], with the photons and electrons swapped. This setup raises two questions. First, does the *"Preferred way"* interpretation consider M as a measuring device? In other words, does the photon from S immediately "feel" the mirror and "perceive" it as a potential detector? Second, does it perceive the electron gun (EG) as a device for position measurement?

If the answer is "Yes" to at least one of these questions, the *"Preferred way"* predicts the *InstReguration* of the state in Fig-s 2, 7 to path SM only, with 100% of detection. Numerically, that would give the result (3.4) for SW for the *whole photon* emitted toward M.

But conventional QM predicts that *maximal* detection rates for the case shown in Fig.7 would be only one half of the result (3.4). The missing half (absence of detections, state $|0\rangle_R$) is the hallmark of the photon's left half-self. It does not InstRegurate to merge with its right half-self when the gun is inserted on the right. All outcomes with $|0\rangle_R$ indicate the InstReguration to $|1\rangle_L$.

Such cases flatly contradict the claimed tendency of the observed entity to satisfy the observer. But they all confirm the *probabilistic InstReguration* from the wave behavior to particle behavior according to the Born rule.

**II**. This can be farther proved by inserting an additional EG on the left side before the arrival of the left half-self (Fig.8). Then whenever shooting on the right misses the prey, the photon will be recorded in its entirety on the *left* side of S.

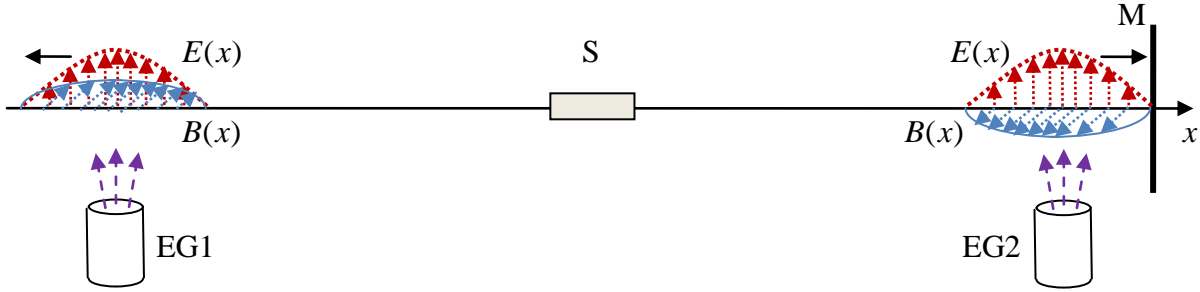

**Fig. 8**
The split photon from S interacts with one of the two EG. The EG locations may also be asymmetric with respect to S, with the corresponding adjustment for the shooting times.

The left EG may give positive outcome even when inserted *after* the right one, or vice versa. The symmetric setup in Fig.8 is only a special case. The left EG can be placed at an arbitrary distance $L$ from S, with corresponding time interval for shooting

$$\frac{L-a/2}{c} \leq t \leq \frac{L+a/2}{c} \qquad (4.7)$$

Assuming 100% efficiency of shooting on either side, each trial will result in the photon detection. The statistics of detections is insensitive to the time ordering of shootings. The photon must be detected on one side of S in 50% of trials regardless of when if ever the other EG is used. If it is used, the missing 50% of detections are destined to happen there.

**III**. Insert a photon detector between M and the approaching right pulse at a distance $S$ from the source, that is, within the time interval

$$0 < t < \frac{S-a/2}{c} \qquad (4.8)$$

If the photon always chooses the path leading to detector, the detector would click after each emission. In reality, it will click only half of the time, according to 50% probability represented by each half-self of the photon.

The same would happen with the detector on the opposite side of S. And with both detectors inserted (Fig.9), they would click in anti-coincidence at each trial regardless of the ordering of their insertion. The statistics must be the same as in case II.

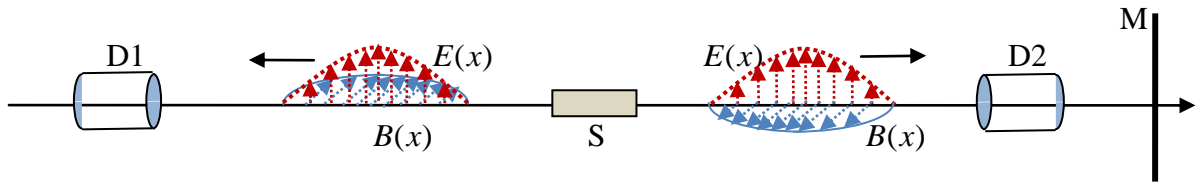

**Fig. 9**

The split photon from S in experiment with delayed insertion of one or two photon detectors D within time interval (4.8).

**IV**. Finally, consider post-reflection stage, but with S removed or kept transparent until the next trial. That will unblock the way back for the reflected half-self of the photon. Both pulses are now moving in one direction, at a constant distance $2D$ from each other. This leads to variety of choices briefly outlined below.

**(a)** Use only one detector and insert it behind the reflected part (Fig.10). There will be no detections. It's too late. Our world is governed by retarded causality.

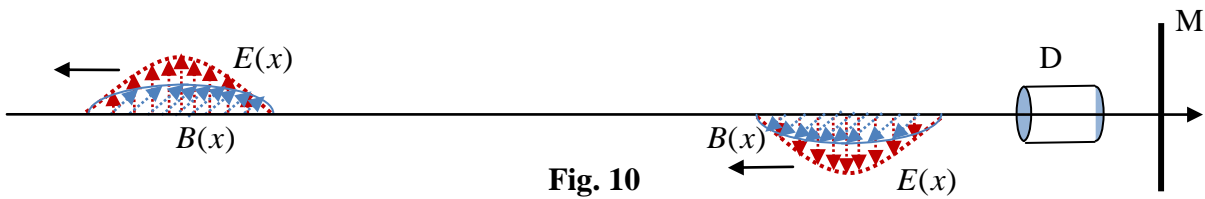

**Fig. 10**

**(b)** Insert detector between the pulses (Fig.11). The leading half-self remains beyond the reach of detector. The detector must click in half of the trials at the time of arrival of reflected half-self. All positive outcomes show the InstReguration of the split photon entirely to its trailing half-self. Quantum non-locality works backward in such events.

The trials without clicks show the photon's choice to InstRegurate into its leading part to remain undetected. The non-locality now works forward, precluding intended observation. There is no preferred choice in such statistics.

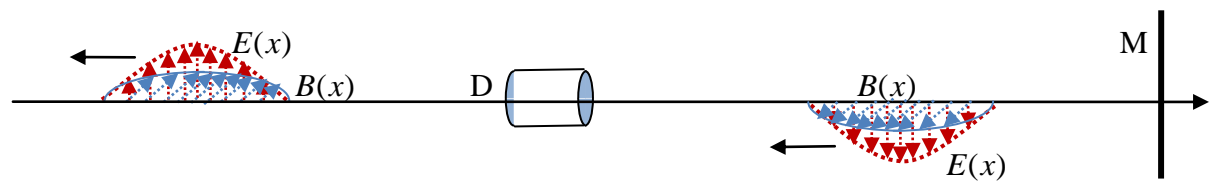

**Fig. 11**

**(c)** Place the detector in front of both pulses (Fig.12). The photon will be recorded in all trials, either at the time of arrival of its leading half-self (non-locality working forward), or later at the time of arrival of trailing half-self (non-locality working backward).

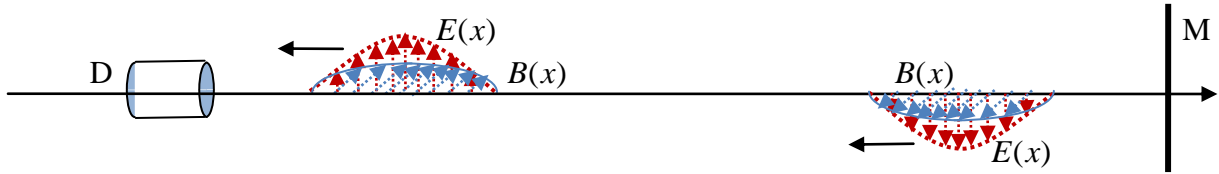

**Fig. 12**

**(d)** Place two detectors between the pulses (Fig.13). According to the "Preferred way" interpretation, all trials must be successful once there is at least one detector in front of the reflected half-self. According to conventional QM, the photon will be detected only in half of the trials – by detector closer to the mirror, at the arrival time of the reflected pulse. Non-locality works backward in such outcomes and forward in cases of non-detection.

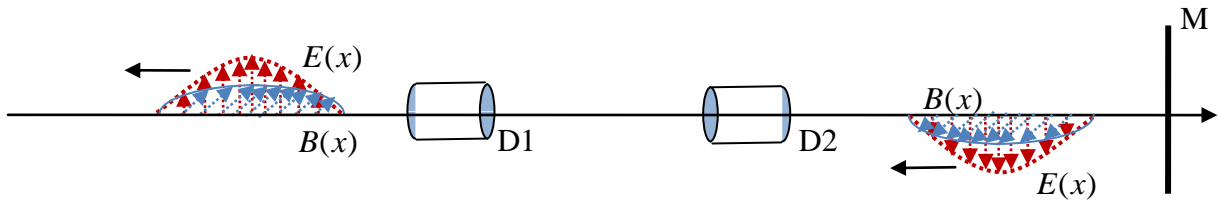

**Fig. 13**

**(e)** Place D2 between the pulses, and D1 – in front of both (Fig.14). The photon must be recorded in all trials. Half of them will be from D1 at the time of arrival of the leading half-self (non-locality working forward). Another half will be done by D2 at the moment of arrival of the trailing half-self (non-locality working backward).

The "Preferred way" interpretation does not specify, whether the photon must InstRegurate into the pulse approaching the detector inserted earlier or to detector that is closer even if inserted later. Therefore it cannot be tested in this case.

According to conventional QM, either detector will remain idle in half of the trials. Denoting an idle or clicking state of a detector as $|0\rangle_D$ and $|1\rangle_D$, respectively, we can write the two possibilities as $|0\rangle_{D1}|1\rangle_{D2}$ or $|1\rangle_{D1}|0\rangle_{D2}$. Each possibility has 50% chance to materialize, albeit at different moments of time.

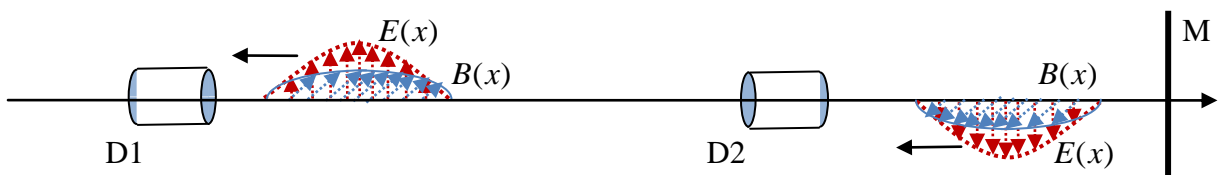

**Fig. 14**

**(f)** Place both detectors in front of the pulses (Fig.15). QM predicts the same outcomes as in the previous case, with the same chances to actualize.

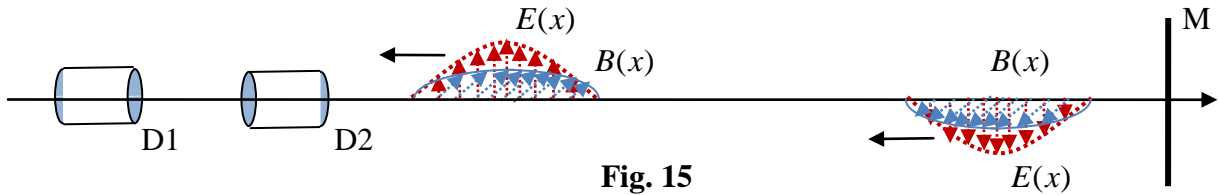

**Fig. 15**

All predictions of QM for considered cases are in agreement with each other and refute the "*Preferred way*" interpretation. One might argue that this interpretation does not apply to signals because they do not feel distant objects beyond their edges and therefore cannot instantly select the "right way" to detector. But the experiments with smoothly shaped wave forms had not confirmed "*Preferred way*" interpretation either.

The discussed experimental setup allows more varieties, for instance, if we insert additional mirror to the left of the source. But the analysis will lead to the same predictions confirming conventional QM.

## Conclusions

An interesting feature of signal reflection is the appearance of discontinuities $\mathcal{D}$ in its inner region during the reflection process. In case of a photon, both $\mathcal{D}$ are in the field's derivatives, apart from the initial discontinuity in the *E*-field's derivative and the newly-formed discontinuity in the *B*-field itself, both remaining at the reflecting surface. The inner $\mathcal{D}$-s coincide and move together – first away from and then toward the mirror.

The outer discontinuities (the edges in the signal form) allow one to quantify the respective non-locality, which provides new guidance – the appropriate time schedule for the respective delayed choice experiments.

In all cases, the detector remaining idle after the moment of pulse arrival indicates the existence of the other part(s) of the packet in some other region(s) of space. So the photon does not exclusively select the one (shortest) way leading to detector – it takes all available paths, and the part reaching the detector interacts with it probabilistically, due to existence of the other even if very remote part(s). This reflects the intrinsically probabilistic nature of a split signal.

The basic feature of the considered state is the existence of non-local connection between its separate parts *despite the zero field amplitude in all separating space*. Due to this inextricable non-locality, each half-self of the photon must immediately feel the other half-self's engagement with detector.

The delayed insertion of a detector does not retroactively change the statistics of observations. This is consistent with the results from the previously studied versions of the experiment [6-19]. The corresponding outcomes are just another demonstration of the wave-particle duality. This again illustrates the basic tenet of QM: a QM object is neither exactly wave, nor exactly particle [28, 29], but it can show either of these faces of reality in a respective measurement.

# References

1. Moses Fayngold, Vadim Fayngold,
   *Quantum Mechanics and Quantum Information*, Wiley-VCH, 2013
2. M. Fayngold, Resonant Scattering and Suppression of Inelastic Channels in a Two-Dimensional Crystal, *Sov. Phys. JETP*, **52**, 408 (1985)
3. M. Fayngold, On the Algebraic Structure of Elastic Scattering Amplitude in a Two-Dimensional Crystal, *Sov. Phys. Journ*. No.4, 30 (1987)
4. S. Lebegue, O, Eriksson,
   Electronic structure of two-dimensional crystals from ab initio theory,
   *Phys. Rev.* B **79** (11), 2009
5. A. H. Castro Neto, K. Novoselov, Beyond Graphene, *Materials Express* **1** (1), March 2011
6. J. A. Wheeler, in *Mathematical Foundations of Quantum Mechanics*, Ed. A.R.Marlow, Academic Press, New York, pp. 9-48, 1978
7. Wheeler, J. A. in *Quantum Theory and Measurement,*
   (Ed-s Wheeler, J. A. & Zurek, W. H.), 182–213 (Princeton University Press, 1984)
8. T. Hellmuth, H. Walther, A. Zajonc and W. Schleich,
   Delayed-choice experiments in quantum interference, *Phys. Rev*. A, **35**, 2532-2541 (1987)
9. C. O. Alley, O. Jakubowicz, C. A. Steggerda and W. C. Wickes,
   A delayed random choice quantum mechanic experiment with light quanta,
   *Proceedings of the International Symposium on the Foundations of Quantum Mechanics*,
   Ed. S. Kamefuchi (Phys. Soc. of Japan), pp. 158-164, 1983
10. Nick Herbert, *Quantum Reality*, Anchor Books, Doubleday, pp. 164-166, 1987
11. G. Greenstein, A. G. Zajonc, *The Quantum Challenge, Modern Research on the Foundations of Quantum Mechanics*, Jones and Bartlett Publishers, 1997
12. Alberto Peruzzo, Peter Shadbolt, Nicolas Brunner, Sandu Popescu, and Jeremy L. O'Brien, Quantum delayed choice experiment, arXiv:1205.4926v2 [quant-ph] 28 Jun 2012
13. Lawson-Daku, B. J. et al. Delayed choices in atom Stern–Gerlach interferometry,
    *Phys. Rev. A* **54**, 5042–5047 (1996)

14. Y. H. Kim, R. Yu, S. P. Kulik, Y. Shih, and M. O. Scully, Delayed 'choice' quantum eraser,
*Phys. Rev. Let.* **84**, 1–5 (2000)

15. T. Hellmut, H. Walther, A. G. Zajonc, & W. Schleich,
    Delayed-choice experiments in quantum interference, *Phys. Rev. A* **35**, 2532–2541 (1987)
16. J. Baldzuhn, E. Mohler, & W. Martienssen,
    A wave–particle delayed-choice experiment with a single-photon state,
    *Z. Phys. B* **77**, 347–352 (1989)
17. V. Jacques, et al. Experimental realization of Wheeler's delayed-choice gedanken Experiment, *Science* **315**, 966–968 (2007)
18. V. Jacques, et al. Delayed-choice test of quantum complementarity with interfering single photons, *Phys. Rev. Let.* **100**, 220402 (2008)
19. R. Ionicioiu, & D. R. Terno, Proposal for a quantum delayed-choice experiment,
    *Phys. Rev. Let.* **107**, 230406 (2011)
20. G. Gamow, "Zur quantentheorie de atomkernes," *Z. Phys*. **51**, 204–212 (1928)
21. Ya. B. Zeldovich, "On the theory of unstable states," *Sov. Phys. JETP* **12**, 542–548 (1961)
22. R. M. More, "Theory of decaying states," *Phys. Rev*. A, **4**, 1782–1790 (1973)
18